\documentstyle[aps,pre,epsf,twocolumn]{revtex}

\begin{document}
\draft

\twocolumn[
\hsize\textwidth\columnwidth\hsize\csname @twocolumnfalse\endcsname

\title{Surface critical behavior of driven diffusive systems with open
boundaries}
\author{K. Oerding and H. K. Janssen}
\address{
Institut f\"ur Theoretische Physik III\\
Heinrich-Heine-Universit\"at\\
D-40225 D\"usseldorf, Germany
}
\date{\today}
\maketitle

\widetext
\begin{abstract}
Using field theoretic renormalization group methods we study the critical
behavior of a driven diffusive system near a boundary perpendicular to
the driving force. The boundary acts as a particle reservoir which is
necessary to maintain the critical particle density in the bulk.
The scaling behavior of correlation and response functions is governed by
a new exponent $\eta_{1}$ which is related to the anomalous scaling
dimension of the chemical potential of the boundary.
The new exponent and a universal amplitude ratio for the density profile
are calculated at first order in $\epsilon = 5-d$.
Some of our results are checked by computer simulations.
\end{abstract}

\narrowtext
]

\section{Introduction}

In order to study the properties of thermodynamic
systems far from equilibrium physicists have been looking for simple
models which capture the main features of non-equilibrium phenomena.
Driven diffusive systems (DDS) introduced by Katz et. al.~\cite{KLS} to
model fast ionic conductors are characterized by a particles conserving
dynamics and a stationary state which does not satisfy detailed balance.
Their study has led to the discovery of the connection between
the validity of a conservation law and the existence of long--range
spatial correlations in non--equilibrium steady states.

A simple microscopic realization of DDS is an Ising lattice gas with
attractive nearest neighbor interaction and an external driving force
$E$ which prefers particle jumps in the direction parallel to
$E$~\cite{KLS,K-tL}. The strength of the particle attraction can
be varied by a temperature--like parameter $T$. Below a critical
value $T_{c}(E)$ particles are separated in the stationary state into
regions of high and low densities, where the interfaces are oriented
parallel to the driving force (for $E \neq 0$). The phase transition
at $T_{c}(E)$ is second order. For an infinite driving force
particle jumps in the direction antiparallel to $E$ are suppressed.
In this case the phase transition occurs at $T_{c}(\infty) \approx 1.41
\,T_{c}(0)$, where $T_{c}(0)$ is the critical temperature of the two
dimensional Ising model~\cite{K-tL}.
The critical behavior of this system has been extensively studied by
Monte Carlo simulations and renormalization group methods~\cite{JS}.
(For a recent review see~\cite{SZ}.)

In most studies of DDS periodic boundary conditions in all directions
are imposed to avoid surface effects. In this case the particles are
driven along a ring or torus. In more realistic models particles are
fed into the system at one side and extracted at the other.
The asymmetric exclusion model~\cite{DDM} is a DDS with hard core
repulsion (but without nearest neighbor interaction, i.e. $T = \infty$).
The density profile in a one-dimensional exclusion model
with a particle source and a sink was investigated numerically by
Krug~\cite{JK}. His results were later confirmed and generalized by
exact calculations~\cite{SD,DEHP,Fab}.
An important result of these works is the `maximum current principle'
which states the following:  If the system is placed between two particle reservoirs A and B (with the respective densities $c_{A}$ and $c_{B}$
with $c_{A} \geq c_{B}$) and the driving force points from
A to B, then the bulk density takes the value $c_{\text{max}}$
which maximizes the current $j$ under the constraint $c_{A} \geq c
\geq c_{B}$, i.e.
\begin{equation}
j(c_{\text{max}}) = \max\{ j(c) \mid c_{A} \geq c \geq c_{B} \}.
\end{equation}
The maximum current density of an Ising lattice gas with attractive
particle-particle interaction equals its critical density $1/2$.
The low temperature phase of this system 
with open boundaries in two dimensions has been studied by Boal et.
al.~\cite{BSZ}.

In the present paper field theoretic renormalization group methods are
employed to investigate the effects of open boundaries on DDS at the
critical point $T_{c}(E)$.
We assume that a plane particle source A perpendicular to the
driving force is located at the left boundary of the system (coordinate
$z=0$) and impose periodic boundary conditions in the transverse
directions. The effect of the particle source is to suppress density
fluctuations and to maintain a constant density $c_{A}$ at $z=0$.
The particles are extracted from the system when they reach a sink B
located at $z=L$ ($L \to \infty$).

It is well known that in physical systems with long ranged correlations
the influence of a surface extends far into the bulk. The critical
behavior near a boundary is governed by universal scaling laws with
critical exponents that (in general) cannot be expressed in terms of
bulk exponents (a review is given in Ref.~\cite{Diehl}).
The applicability of renormalization group methods to investigate
both static~\cite{Diehl,Diet,Diet2,Diet3,Sym} and
dynamic~\cite{DD83,DJ,Diehl2} surface universality classes
is well established. Especially encouraging are the results of
Ref.~\cite{JO} where this technique has been used to obtain an
approximate profile for one-dimensional DDS with open boundaries.
It turned out that the profile calculated by renormalization group
improved perturbation theory (at one-loop order) was in good agreement
with the exact result of Ref.~\cite{SD}.


In the next section the semi-infinite extension of the field theoretic
model for DDS at the critical point (introduced in~\cite{JS}) is
presented. Above the upper critical dimension $d_{c}=5$ of this model
fluctuations around the mean field profile can be treated by na{\"\i}ve
perturbation theory. The mean field profile and Gaussian fluctuations
for $d>5$ are considered in some detail in Secs.~\ref{mean-field}
and~\ref{d>5} since the results of this analysis remain qualitatively
valid for $d<5$. In Sec.~\ref{RGanal} the renormalization group is
used to obtain the scaling behavior of Green functions and the density
profile below five dimensions.
Sec.~\ref{Disc} contains a discussion.

\section{The model}

The analysis in the present paper is based on the field theoretic model
introduced by Janssen and Schmittmann~\cite{JS} to study the critical
behavior of a diffusive system with a single conserved density subjected
to an external driving force. The model can be written in the form of
the continuity equation
\begin{equation}
\partial_{t} s + \nabla_{\bot} {\bf j}_{\bot} + \partial_{\|} j_{\|} = 0 ,
\end{equation}
where $s({\bf r}, t) = c({\bf r}, t) - \bar{c}$ denotes the deviation of
the concentration from its average (bulk) value $\bar{c}$, and ${\bf
j}_{\bot}$ and $j_{\|}$ are the respective components of the current
perpendicular and parallel to the driving force.  
The explicit expression for the current can be motivated by the following
symmetry requirements~\cite{SZ}:
\begin{itemize}
\item[(i)]
Isotropy with respect to the $(d-1)$-dimensional transverse subspace,
\item[(ii)]
invariance of the equation of motion under reversal of the driving
force ($E \leftrightarrow -E$) and particle-hole exchange
(`charge conjugation', $s \leftrightarrow -s$),
\item[(iii)]
invariance under force reversal and reflection in $r_{\|}$ (the
coordinate parallel to the force).
\end{itemize}
A continuum model satisfying (i)--(iii) describes, for instance, the
long time and large distance behavior of a driven Ising lattice gas at
its critical density $1/2$. Since the current in this system is at its
maximum value for half filling one may use the maximum current principle
(in a system with open boundaries) to adjust the bulk density to the
critical value.

Keeping only terms which are consistent with the above
conditions~(i)--(iii) and relevant or marginal in the renormalization
group sense the current may be written (upon rescaling of $s$) in the
form
\begin{mathletters}
\begin{eqnarray}
j_{\bot} = \nabla_{\bot} \left[-\lambda \left( \tau s - \triangle_{\bot} s
\right) + \zeta \right] \label{ja} , \\
j_{\|} = E \left(\sigma_{0} + \sigma_{2} s^{2}\right) - \lambda \rho
\partial_{\|} s , \label{jb}
\end{eqnarray}
where $\zeta$ is a Gaussian random force with the correlation
\begin{equation}
\langle \zeta({\bf r}, t) \zeta({\bf r}^{\prime}, t^{\prime}) \rangle =
2 \lambda \delta({\bf r} - {\bf r}^{\prime}) \delta(t-t^{\prime}) .
\end{equation}
\end{mathletters}
The third order derivative in Eq.~(\ref{ja}) has to be kept because the
coefficient $\tau$ of the first order derivative vanishes at the critical
point (transverse phase transition~\cite{JS}).
The coefficient $\sigma_{0}$ may be interpreted as the conductivity of
the system at the maximum current density $\bar{c}$. Deviations from
$\bar{c}$ due to fluctuations or a non-constant density profile decrease
the current. This effect is modeled by the term $E\sigma_{2}s^{2}$ in
Eq.~(\ref{jb}). The coefficient $\tau$ measures the deviation of the
temperature parameter from its critical value and $\rho$ takes into
account the anisotropy of the diffusion constant. Even if the diffusion
constant is isotropic ($\rho=1$) in the original Langevin equation it
becomes anisotropic under coarse graining.

For the subsequent field theoretic analysis it is convenient to recast the
model in the form of the dynamic functional~\cite{J76,DD,BJW,DP,J92,JS}
\begin{eqnarray}
{\cal J}_{\text{b}}[\tilde{s}, s]
=& \int dt \int_{V} d^{d}r \Bigl\{ \tilde{s} \partial_{t} s
+ \lambda \bigl[(\triangle_{\bot} \tilde{s}) (\triangle_{\bot} s) \nonumber \\
&+ \tau (\nabla_{\bot} \tilde{s}) (\nabla_{\bot} s)
+ \rho (\partial_{\|} \tilde{s}) (\partial_{\|} s) \label{Jb} \\
&+ \frac{1}{2} g (\partial_{\|} \tilde{s}) s^{2} - h \partial_{\|} \tilde{s}
- (\nabla_{\bot} \tilde{s})^{2} \bigr]\Bigr\} , \nonumber
\end{eqnarray}
where $\lambda g \sim -E \sigma_{2}$, and $\tilde{s}$ is a
Martin--Siggia--Rose response field~\cite{MSR}.
While the functional~(\ref{Jb}) allows us to calculate response and correlation
functions for an infinite bulk system the influence of the boundaries has to
be modeled by additional surface contributions.
If the boundary is perpendicular to the driving force the region of integration
in Eq.~(\ref{Jb}) is the half space $V=\{({\bf r}_{\bot}, z) \mid {\bf
r}_{\bot}\in {\bf R}^{d-1}, z\geq 0 \}$, and the surface is defined by $z=0$.
Omitting irrelevant and redundant terms~\cite{JO} the surface functional
reads
\begin{eqnarray}
{\cal J}_{\text{s}}[\tilde{s}, s] =& \int dt \int_{\partial V}
d^{d-1}r_{\bot} \lambda \Bigl( c \tilde{s} s - \tilde{c} \tilde{s}^{2} -
c_{2} \tilde{s}
\triangle_{\bot} s \nonumber \\
&+ \frac{1}{2} g_{s} \tilde{s} s^{2} - h_{s} \tilde{s} \Bigr)
\end{eqnarray}
Response and correlation functions can now be calculated
by functional integration with the weight
$\exp(-{\cal J})$, where ${\cal J} = {\cal J}_{\text{b}} +
{\cal J}_{\text{s}}$.

\section{Equation of motion and mean field approximation}
\label{mean-field}

An exact equation for the stationary profile $\Phi(z) =
\langle s({\bf r}, t) \rangle$ follows from the invariance of the
generating functional
\begin{eqnarray}
Z[\tilde{J}, J; \tilde{J}_{1}, J_{1}] = \int {\cal D}[\tilde{s}, s]
\exp\Bigl( -{\cal J}_{b}[\tilde{s}, s] - {\cal J}_{1}[\tilde{s}_{s}, s_{s}]
\nonumber \\
 + (\tilde{J}, \tilde{s}) + (J, s) + (\tilde{J}_{1},
\tilde{s}_{s}) + (J_{1}, s_{s}) \Bigr) \label{genfun}
\end{eqnarray}
under an infinitesimal shift of the field $\tilde{s}$. In Eq.~(\ref{genfun})
$s_{s}({\bf r}_{\bot}) = s({\bf r}_{\bot}, 0)$
denotes the field at the surface and the abbreviations
\begin{equation}
(\tilde{J}, \tilde{s}) = \int dt \int_{V} d^{d}r \tilde{J} \tilde{s}
\end{equation}
and
\begin{equation}
(J_{1}, s_{s}) = \int dt \int_{\partial V} d^{d-1}r_{\bot} J_{1} s_{s}
\end{equation}
have been used. The invariance of $Z[\tilde{J}, J; \tilde{J}_{1}, J_{1}]$
implies the equation of motion
\begin{equation}
\partial_{t} s + \lambda \bigl[ (\triangle_{\bot} - \tau
) \triangle_{\bot} s - \partial_{\|} (\rho \partial_{\|} s + \frac{1}{2}
g s^{2} ) + 2 \triangle_{\bot} \tilde{s} \bigr] = \tilde{J} \label{eqmo}
\end{equation}
which holds after insertion into averages.
The invariance of $Z[\tilde{J}, J; \tilde{J}_{1}, J_{1}]$ under a shift
of the surface field $\tilde{s}_{s}$ leads to the equation of motion
\begin{eqnarray}
-\rho \partial_{n} s - \frac{g - g_{s}}{2} s_{s}^{2} + c s_{s}
- 2 \tilde{c} \tilde{s}_{s} \nonumber \\
- c_{2} \triangle_{\bot} s_{s} - h_{s} + h = \frac{1}{\lambda}
\tilde{J}_{1} \label{eqmos}
\end{eqnarray}
which fixes the boundary condition.
Taking the average on both sides of~(\ref{eqmo}) for vanishing sources
$\tilde{J}=J=\tilde{J}_{1}=J_{1}=0$ one obtains
\begin{equation}
\partial_{z} \bigl[ \rho \partial_{z} \Phi(z) + \frac{1}{2}
g \bigl( \Phi(z)^{2} + C(z) \bigr) \bigr] = 0
\end{equation}
or, since $\Phi_{\text{bulk}}=0$ due to the definition of $s$,
\begin{equation}
\Phi^{\prime}(z) + \frac{g}{2 \rho}
\bigl( \Phi(z)^{2} + C(z) - C_{\text{bulk}} \bigr) = 0 . \label{eqphi}
\end{equation}
The function
$C(z) = \langle [ s({\bf r}_{\bot}, z; t) - \Phi(z) ]^{2} \rangle$
describes density fluctuations at the distance $z$ from the surface
and $C_{\text{bulk}}$ denotes its value for $z \to \infty$.

In the mean field approximation one neglects the correlation function
$C(z)$ and obtains for the profile
\begin{equation}
\Phi_{\text{mf}}(z) = \Phi_{0} \Bigr( 1 + \frac{g}{2 \rho} \Phi_{0} z
\Bigl)^{-1} . \label{phimf}
\end{equation}
Dimensional analysis shows that the momentum dimension of the coupling
coefficient $g$ is given by $[g] = (5-d)/2$, and the mean field
approximation breaks down below the upper critical dimension $d_{c}=5$.
For $d>5$ corrections to the mean field profile can be obtained by
na{\"\i}ve perturbation theory. At lowest order it is sufficient to
calculate the perturbation $C(z)-C_{\text{bulk}}$ in Eq.~(\ref{eqphi})
by a Gaussian approximation.

\section{Corrections to the mean field profile for $d>5$}
\label{d>5}

In the simplest case, $\Phi_{\text{mf}}(z) =
\Phi_{0}=h_{s}=0$, the Fourier transform (with respect to
${\bf r}_{\bot}$ and $t$) of the Gaussian propagator
\begin{equation}
G({\bf r}_{\bot}, z, z^{\prime}; t) = \langle s({\bf r}_{\bot},
z; t) \tilde{s}({\bf 0}, z^{\prime}; 0) \rangle_{0}
\end{equation}
is given by~\cite{JO}
\begin{eqnarray}
\hat{G}_{{\bf q}_{\bot}, \omega}(z, z^{\prime})
= \frac{1}{2 \lambda \sqrt{\rho} \kappa} \Bigl[ e^{-\kappa
|z-z^{\prime}|/\sqrt{\rho}} \nonumber \\
+ \frac{\kappa - c/\sqrt{\rho}}{\kappa + c/\sqrt{\rho}} e^{-\kappa
(z+z^{\prime})/\sqrt{\rho}} \Bigr] \label{prop}
\end{eqnarray}
with
\begin{equation}
\kappa = \Bigl( \frac{i \omega}{\lambda} + q_{\bot}^{2} (\tau + 
q_{\bot}^{2})\Bigr)^{1/2} .
\end{equation}
The parameter $c$ occurring in the surface functional ${\cal J}_{1}$ and
in the propagator describes (for $c>0$) the suppression of density
fluctuations by the particle reservoir at the boundary. Since the
momentum dimension of $c$ is one the asymptotic scaling behavior is
governed by the fixed point $c_{\star}=\infty$. At this fixed point the
fields $\tilde{s}$ and $s$ satisfy the Dirichlet boundary conditions
$\tilde{s}_{s}=s_{s}=0$.

The Fourier transform of the Gaussian correlation function
\begin{equation}
C({\bf r}_{\bot}, z, z^{\prime}; t) = \langle s({\bf r}_{\bot},
z; t) s({\bf 0}, z^{\prime}; 0) \rangle_{0}
\end{equation}
at the Dirichlet fixed point $c_{\star}=\infty$ can be derived from the
Gaussian part of the dynamic functional. This calculation yields
\begin{eqnarray}
\hat{C}_{{\bf q}_{\bot}, \omega}(z, z^{\prime}) & = &
2 \lambda q_{\bot}^{2} \int_{0}^{\infty} d z^{\prime\prime}
\hat{G}_{{\bf q}_{\bot}, \omega}(z, z^{\prime\prime})
\hat{G}_{{\bf q}_{\bot}, -\omega}(z^{\prime\prime}, z^{\prime})
\nonumber \\
& = & -\frac{2 \lambda q_{\bot}^{2}}{\omega} \Im [
\hat{G}_{{\bf q}_{\bot}, \omega}(z, z^{\prime}) ] , \label{corr}
\end{eqnarray}
where $\Im[\cdots]$ denotes the imaginary part.
The equal-time correlation function at the point $({\bf r}_{\bot}, z)$
is given by
\begin{eqnarray}
C(z) & = & \int_{{\bf q}_{\bot}, \omega} \hat{C}_{{\bf q}_{\bot},
\omega}(z, z) \nonumber \\
& = & C_{\text{bulk}} - \frac{1}{2 \sqrt{\rho}} (8 \pi
z/\sqrt{\rho})^{-(d-1)/2} \label{Cz}
\end{eqnarray}
with $C_{\text{bulk}} \sim \Lambda^{d-1}/\sqrt{\rho}$ (where $\Lambda$
is a cut-off wave number).

We can now use Eqs.~(\ref{eqphi}) and~(\ref{Cz}) to compute the
fluctuation correction to the constant mean field profile
$\Phi_{\text{mf}}(z)=0$. At first order in $g$ we get
\begin{equation}
\Phi^{[1]}(z) = - \frac{g (8 \pi)^{-(d-1)/2}}{2 (d-3) \rho} \Bigl(
\frac{z}{\sqrt{\rho}} \Bigr)^{-(d-3)/2} . \label{Phi1}
\end{equation}
One can easily check by dimensional analysis that higher order corrections
to the profile decay as $\Phi^{[2n+1]}(z)/\Phi^{[1]}(z) \sim z^{-n (d-5)/2}$
(up to cut-off dependent terms which may change the amplitude of the
leading term proportional to $z^{-(d-3)/2}$).

\begin{figure}
\epsfxsize=300pt
\vspace*{-48mm}\hspace*{-13mm}
\epsfbox{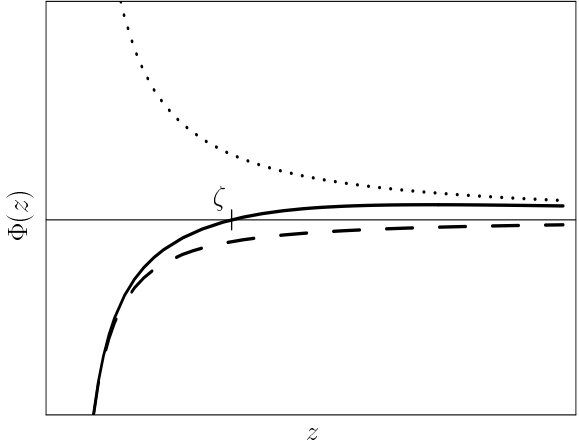}
\vspace*{-40mm}
\caption{Sketch of the profile $\protect\Phi(z)$ for $h_{1}
= \protect\infty$ (dotted), $h_{1} > 0$ finite (solid curve), 
and $h_{1} = 0$ (dashed). For $h_{1} < 0$ the density
in the bulk stays below its critical value indicated by the horizontal line.}
\label{sketch}
\end{figure}

In the limit $c, h_{s} \to \infty$ (with $h_{s}/c =: h_{1}$ fixed)
the boundary value of the mean field profile is given by $\Phi_{0} =
h_{1}$. This follows from Eqs.~(\ref{eqmos}) and~(\ref{phimf}).
For $\Phi_{0} > 0$ the mean field profile decays asymptotically as
$\Phi_{\text{mf}}(z) \simeq 2 \rho/(g z)$, and the fluctuation correction
$\sim z^{-(d-3)/2}$ can be neglected for $z \to \infty$ ($d>5$). However,
if $h_{1}$ is positive but small the profile $\Phi(z)$ is negative
for $z < \zeta$, where $\zeta = \zeta(h_{1})$ is a crossover length.
The dependence of the profile on the boundary chemical potential
$h_{1}$ is depicted qualitatively in Fig.~\ref{sketch}. 
The crossover length $\zeta$ tends to infinity for $h_{1} \to 0^{+}$.
To estimate $\zeta$ for small $h_{1}$ we equate the mean field
profile $\Phi_{\text{mf}}(\zeta)$ with the fluctuation term $\zeta^{-(d-3)/2}$
and obtain
\begin{equation}
\zeta \sim h_{1}^{-2/(d-3)}
\end{equation}
for $h_{1} \to 0^{+}$, $d > 5$.

In the language of semi-infinite magnetic systems the case $h_{1} = \infty$ corresponds to the normal transition and $h_{1} = 0$ is the ordinary point.
A length scale similar to $\zeta$ has already been found in magnetic
systems near the ordinary transition~\cite{RC,ACUR}. There the length scale
$\zeta$ characterizes the magnetization profile induced by a
small magnetic surface field. 

\section{Renormalization group analysis}
\label{RGanal}

\subsection{Renormalization}
\label{renorm}

The na\"{\i}ve perturbation theory described in the previous section breaks
down below the upper critical dimension $d_{c}=5$. The renormalization group
allows us to improve the perturbation expansion by a partial resummation.

Since the individual terms of the perturbation series contain for $d=d_{c}$
ultraviolet divergent integrals a regularization prescription is required
to obtain well-defined expressions for the otherwise infinite integrals.
Here we use the dimensional regularization method (analytic continuation of
the integrals as functions of $d$).
The remaining poles in $\epsilon=d_{c}-d$ are then absorbed into
reparametrizations of the coupling coefficients and the fields.
In the field theory for the bulk model (without a surface) a
renormalization of the parameter $\rho = Z \rho_{R}$ is sufficient
to cancel the ultraviolet divergences at every order of the perturbation
theory~\cite{JS}.
At one-loop order the renormalization factor is given by
\begin{equation}
Z = 1-\frac{u}{\epsilon} + O(u^{2}) , \qquad
u = A_{\epsilon} g^{2} \rho_{R}^{-3/2} \mu^{-\epsilon} ,
\end{equation}
where $\mu$ is an external momentum scale and the geometrical factor
$A_{\epsilon} = (3/4)(4\pi)^{-d/2} \Gamma((3-\epsilon)/2)
\Gamma(1+\epsilon/2)/\Gamma(2-\epsilon/2)$ has been introduced for
convenience. (The index `$R$' indicates
renormalized quantities.)

In order to investigate the scaling behavior of response
functions near the boundary one has to calculate Green functions with
insertions of the surface response field $\tilde{s}_{s}$. Since the
Gaussian propagator~(\ref{prop}) vanishes at the Dirichlet fixed point 
$c_{\star}=\infty$ it is necessary to go to higher orders in $c^{-1}$
to obtain non-trivial results~\cite{Diehl,Diet,DDE}. At first order in
$c^{-1}$ the propagator becomes (for $z^{\prime}=0$)
\begin{equation}
\hat{G}_{{\bf q}_{\bot}, \omega}(z, 0) = \frac{1}{c} \left. \rho
\partial_{z^{\prime}} \hat{G}_{{\bf q}_{\bot}, \omega}^{(D)}(z,
z^{\prime}) \right|_{z^{\prime}=0} + \ldots ,
\end{equation}
where $\hat{G}_{{\bf q}_{\bot}, \omega}^{(D)}(z, z^{\prime})$ denotes
the propagator~(\ref{prop}) for $c=\infty$. This shows that the leading
order terms in an expansion in powers of $c^{-1}$ can be studied in the
framework of a field theory with Dirichlet boundary conditions after
replacing in expectation values 
\begin{equation}
\tilde{s}_{s} \rightarrow c^{-1} \rho \partial_{n} \tilde{s} .
\end{equation}
Analogously insertions of the surface field $s_{s}$ have to be
replaced (at leading order) by the the normal derivative $c^{-1}
\rho \partial_{n} s$.

Since a boundary breaks the translational invariance of the system it
gives rise to new divergences in the perturbation series which are
located at the surface [i.e., proportional to $\delta(z)$)]. These
surface divergences have to be subtracted by appropriate counter terms
added to the dynamic functional ${\cal J}$. In the appendix it is shown
that the required counter terms have the form
\begin{eqnarray}
{\cal J}_{\text{bct}}[\tilde{s}, s] = \int dt \int_{V} d^{d}r \lambda
\bigl[\rho_{R} (Z-1) (\partial_{\|} \tilde{s})(\partial_{\|} s)
\nonumber \\
+ \rho_{R}^{-1/2} A_{\epsilon} g \mu^{-\epsilon} A \tau^{2}
\partial_{\|} \tilde{s} \bigr] \label{bct}
\end{eqnarray}
to remove bulk divergences and
\begin{eqnarray}
{\cal J}_{\text{sct}}[\tilde{s}, s] = \int dt \int_{\partial V}
d^{d-1}r_{\bot} \lambda \bigl[ \rho_{R}^{-1/2} A_{\epsilon} g
\mu^{-\epsilon} K (\rho \partial_{n}^{2} \tilde{s}) \hspace*{-4.2pt}
\nonumber \\
+ B (\rho \partial_{n} \tilde{s}) s_{s} + \rho_{R}^{-1} A_{\epsilon} g
\mu^{-\epsilon} F \tau (\rho \partial_{n} \tilde{s})\bigr] \label{sct}
\end{eqnarray}
to cancel $\epsilon$-poles located at the surface. The renormalization
parameters $A$, $B$, $F$, $K$ are calculated at one-loop order with the
result
\begin{equation}
A = -\frac{1}{\epsilon} \qquad B = -\frac{u}{3 \epsilon} \qquad
F = -\frac{4}{3 \epsilon} \qquad K = \frac{2}{3 \epsilon} .
\end{equation}
The first term in ${\cal J}_{\text{bct}}$ renormalizes the diffusion
coefficient $\rho$. The bulk counter term proportional to
$\partial_{\|} \tilde{s}$ corresponds to a renormalization of the
bulk current~\cite{footnote}
\begin{equation}
h = h_{R} - \rho_{R}^{-1/2} A_{\epsilon} g \mu^{-\epsilon} A \tau^{2} .
\end{equation}

We now show that the surface counter terms proportional to the 
operators $(\rho \partial_{n} \tilde{s}) s_{s}$ and $\rho \partial_{n}^{2} 
\tilde{s}$ can be replaced by a multiplicative renormalization of the
surface response field $\rho \partial_{n} \tilde{s}$.
In order to study single insertions of $(\rho \partial_{n} \tilde{s})
s_{s}$ and $\rho \partial_{n}^{2} \tilde{s}$ in Feynman diagrams we first
connect them to the Gaussian response function $\hat{G}_{{\bf q}_{\bot}, 
\omega}^{(D)}(z, z^{\prime})$, i.e. we calculate the Gaussian expectation
values
\begin{eqnarray}
\left\langle s(z) \tilde{s}(z^{\prime})
\int dt^{\prime} \int d^{d-1}x^{\prime} \lambda (\rho \partial_{n}
\tilde{s}) s_{s} \right\rangle_{0} = 0 , \label{a8} \\
\left\langle s(z) (\rho \partial_{n} \tilde{s})
\int dt^{\prime} \int d^{d-1}x^{\prime} \lambda (\rho \partial_{n}
\tilde{s}) s_{s} \right\rangle_{0} \nonumber \\
= \frac{1}{\lambda} e^{-\kappa z/\sqrt{\rho}}
= \left\langle s(z) (\rho \partial_{n} \tilde{s})
\right\rangle_{0} \label{a9}
\end{eqnarray}
and
\begin{equation}
\left\langle s(z) \int d^{d-1}x^{\prime} \lambda \rho \partial_{n}^{2}
\tilde{s} \right\rangle_{0} = - \delta(z) . \label{a10}
\end{equation}
In Eqs.~(\ref{a9}) and~(\ref{a10}) the vertices with normal derivatives
have to be interpreted as
\begin{mathletters}
\begin{equation}
\lambda (\rho \partial_{n} \tilde{s}) s_{s} = \lim_{z \to 0}
\lambda (\rho \partial_{\|} \tilde{s}(z)) s_{s}(z)
\end{equation}
and
\begin{equation}
\lambda \rho \partial_{n}^{2} \tilde{s} = \lim_{z \to 0}
\lambda \rho \partial_{\|}^{2} \tilde{s}(z),
\end{equation}
\end{mathletters}
respectively, where the limit $z\to \infty$ has to be
taken {\em after} the averages $\langle \cdots \rangle_{0}$ have been
performed~\cite{Diehl,BM}.

Equations~(\ref{a8}) and~(\ref{a9}) show that the counter term
$B \lambda (\rho \partial_{n} \tilde{s}) s_{s}$ has an effect only in
diagrams in which it is connected to the surface response field
$\rho \partial_{n} \tilde{s}$. In Green functions with an insertion of
$\rho \partial_{n} \tilde{s}$ it effectively is replaces
$\rho \partial_{n} \tilde{s}$ with $(1-B)\rho \partial_{n} \tilde{s}$.

\begin{figure}
\begin{center}
\hspace*{-1mm}
\epsfxsize=160pt\epsfbox{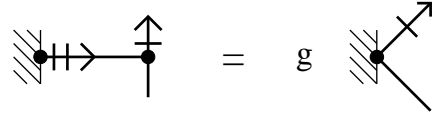}
\vspace*{5mm}
\caption{Effect of the vertex $\protect\lambda \protect\rho 
\protect\partial_{n}^{2} \protect\tilde{s}$ in a Feynman diagram.
The hatched area represents the boundary $z=0$. Each short line
perpendicular to a propagator line indicates a derivative with respect
to $z$.}
\label{f2}
\end{center}
\end{figure}

If the argument of $s$ on the l.h.s. of Eq.~(\ref{a10}) is
fixed (with $z > 0$) the average in~(\ref{a10}) vanishes.
However, due to the $\delta$-function one obtains a non-zero result if
Eq.~(\ref{a10}) is integrated over $z$. Such an integration occurs in
the calculation of the Feynman diagram shown in Fig.~\ref{f2}, where
the counter term vertex proportional to $\lambda \rho \partial_{n}^{2}
\tilde{s}$ is connected to the bulk vertex $g$.
Fig.~\ref{f2} shows that an insertion of the counter term
\begin{displaymath}
\lambda \rho_{R}^{-1/2} A_{\epsilon} g \mu^{-\epsilon} K \rho
\partial_{n}^{2} \tilde{s}
\end{displaymath}
in a Feynman diagram has the same effect
as the vertex $u K \lambda \rho_{R} (\partial_{n} \tilde{s}) s_{s}$.

The above analysis shows that the counter terms proportional to $\rho
\partial_{n}^{2} \tilde{s}$ and $(\rho \partial_{n} \tilde{s}) s_{s}$
give at one-loop order rise to a renormalization of the
surface response field
\begin{equation}
[\rho \partial_{n} \tilde{s}]_{R} = Z_{1}^{-1/2} \rho \partial_{n}
\tilde{s}, 
\end{equation}
where
\begin{equation}
Z_{1}^{-1/2} = 1-B-u K + O(u^{2}) = 1 - \frac{u}{3 \epsilon} + O(u^{2}) .
\end{equation}
The relation between the redundant surface couplings (here $B$, $K$) and
$Z_{1}$ can be extended to higher
orders in $u$ in a similar way as in the case of the
$\phi^{4}$-model~\cite{BM}.

The last counter term in ${\cal J}_{\text{sct}}$ which couples to
$\rho \partial_{n} \tilde{s}$ renormalizes the surface chemical potential
\begin{equation}
h_{1} = Z_{1}^{-1/2} \bigl( [h_{1}]_{R} -
\rho_{R}^{-1} A_{\epsilon} g \mu^{-\epsilon} F \tau \bigr) .
\end{equation}

\subsection{Scaling}
\label{scaling}

With the renormalizations at hand we are in a position to determine
the scaling behavior of the Green functions
\begin{eqnarray}
G_{\tilde{N}, N}^{\tilde{M}, M}(\{ {\bf r}, {\bf x}, t\}) =
\left\langle \prod_{i=1}^{\tilde{N}} \tilde{s}(\tilde{\bf r}_{i},
\tilde{t}_{i}) \prod_{j=1}^{N} s({\bf r}_{j}, t_{j}) \right.
\nonumber \\
\left. \times \prod_{k=1}^{\tilde{M}} \rho \partial_{n}
\tilde{s}(\tilde{\bf x}_{k}, \tilde{t}_{s k})
\prod_{l=1}^{M} \rho \partial_{n} s({\bf x}_{l}, t_{s l})
\right\rangle^{\text{(conn)}} .
\end{eqnarray}
with $\tilde{M}$ insertions of the surface response field
$\rho \partial_{n} \tilde{s}$ and $M$ insertions of $\rho \partial_{n} s$.
In this subsection we omit the index `$R$' since only renormalized
quantities are used.

At the critical point $\tau=h=0$ the Green functions satisfy the
renormalization group equation
\begin{eqnarray}
\Bigl[ \mu \frac{\partial}{\partial \mu} + \beta(u)
\frac{\partial}{\partial u} + \zeta(u) \rho \frac{\partial}{\partial
\rho} + \frac{1}{2} \gamma_{1}(u) h_{1} \frac{\partial}{\partial 
h_{1}} \nonumber \\
+ \frac{\tilde{M}}{2} \gamma_{1}(u) \Bigr] G_{\tilde{N},
N}^{\tilde{M}, M}(\{ {\bf r}, {\bf x}, t\}; u, \rho, h_{1};
\lambda, \mu) = 0 \label{RGE}
\end{eqnarray}
which follows from the independence of the unrenormalized Green functions
from the momentum scale $\mu$. The Wilson functions are given by
\begin{mathletters}
\begin{eqnarray}
\zeta(u) = \left.\mu \frac{d \ln \rho}{d \mu} \right|_{0}
= -u + O(u^{2}) , \\
\beta(u) = \left.\mu \frac{d u}{d \mu} \right|_{0} =
u \Bigl( -\epsilon - \frac{3}{2} \zeta(u) \Bigr) , \\
\gamma_{1}(u) = \left.\mu \frac{d  \ln Z_{1}}{d \mu} \right|_{0} =
-\frac{2}{3} u + O(u^{2}) ,
\end{eqnarray}
\end{mathletters}
where the derivatives are calculated at fixed bare parameters.

The renormalization group equation~(\ref{RGE}) can be solved by
the method of characteristics with the result
\begin{eqnarray}
G_{\tilde{N}, N}^{\tilde{M}, M}(\{ {\bf r}, {\bf x}, t\}; u, \rho,
h_{1}; \lambda, \mu) = X_{1}(l)^{\tilde{M}/2} \nonumber \\
\times G_{\tilde{N}, N}^{\tilde{M}, M}(\{ {\bf r},
{\bf x}, t\}; \bar{u}(l), X(l) \rho, X_{1}(l)^{1/2} h_{1};
\lambda, \mu l) . \label{scal}
\end{eqnarray}
The functions $\bar{u}(l)$, $X(l)$, and $X_{1}(l)$ are solutions of
the set of ordinary differential equations
\begin{mathletters}
\begin{eqnarray}
\frac{d \bar{u}(l)}{d \ln l} = \beta(\bar{u}(l)) , \qquad
\bar{u}(1) = u , \label{flowu} \\
\frac{d \ln X(l)}{d \ln l} = \zeta(\bar{u}(l)) , \qquad X(1) = 1 , \\
\frac{d \ln X_{1}(l)}{d \ln l} = \gamma_{1}(\bar{u}(l)) , \qquad
X_{1}(1) = 1 .
\end{eqnarray}
\end{mathletters}
In the limit $l \to 0$ the scale dependent coupling coefficient
$\bar{u}(l)$ tends to the fixed point value $u_{\star}=(2/3)\epsilon
+ O(\epsilon^{2})$ and the Green functions display power law scaling
with
\begin{eqnarray}
X(l) \simeq X_{\star} l^{-2\eta}, \qquad \eta = -\frac{1}{2}
\zeta(u_{\star}) = \frac{5-d}{3} , \\
X_{1}(l) \simeq X_{1, \star} l^{\eta_{1}} , \qquad \eta_{1} =
\gamma_{1}(u_{\star}) = -\frac{4 \epsilon}{9} + O(\epsilon^{2}) .
\end{eqnarray}
The amplitudes $X_{\star}$ and $X_{1, \star}$ are not universal.

Eq.~(\ref{scal}) can be further simplified by dimensional analysis.
The momentum dimensions of $r_{\bot}$, $r_{\|}$, $t$, $\tilde{s}$,
$s$, and $h_{1}$ follow from the form of the functional
${\cal J}$ (which has to be dimensionless) and are given by
\begin{eqnarray}
r_{\bot} \sim \mu^{-1} , \qquad r_{\|} \sim \rho^{1/2} \mu^{-2} ,
\qquad t \sim \lambda^{-1} \mu^{-4} , \nonumber \\
\tilde{s} \sim \rho^{-1/4} \mu^{(d+3)/2} , \qquad s \sim h_{1}
\sim \rho^{-1/4} \mu^{(d-1)/2} ,
\end{eqnarray}
respectively.
For small $l$ Eq.~(\ref{scal}) maps the large length and time scales
of the critical region on scales on which Green functions can be
calculated perturbatively. Here we are especially interested in the
profile $\Phi(z) = G_{0, 1}^{0, 0}(z)$.
Choosing for the flow parameter the value
\begin{equation}
l = \Bigl( \frac{\mu^{2} z}{\sqrt{\rho X_{\star}}} \Bigr)^{-1/(2+\eta)}
\rightarrow 0 \label{ell}
\end{equation}
we obtain from Eq.~(\ref{scal}) in conjunction with dimensional analysis
the scaling form
\begin{equation}
\Phi(z, h_{1}) = a z^{-\sigma} F\bigl(b
h_{1} z^{1/\nu_{1}} \bigr) , \label{prof}
\end{equation}
with the exponents
\begin{equation}
\sigma = \frac{d-1+\eta}{2 (2+\eta)} = \frac{1+d}{11-d} \label{sig}
\end{equation}
and
\begin{equation}
\frac{1}{\nu_{1}} = \frac{d-1+\eta-\eta_{1}}{2 (2+\eta)} = 1 -
\frac{2 \epsilon}{9} + O(\epsilon^{2}) .
\end{equation}
In~(\ref{prof}) $a$ and $b$ are non-universal scale factors while the
scaling function $F$ is universal.

\subsection{A universal amplitude ratio}

We know from the discussion of the mean field profile and the fluctuation
corrections in section~\ref{d>5} that $\Phi(z, h_{1})$ is finite
and non-zero in both cases $h_{1} = \infty$ and $h_{1} = 0$.
It therefore makes sense to define the universal amplitude ratio
\begin{equation}
D = \lim_{z \to \infty} \frac{\Phi(z, 0)}{\Phi(z, \infty)} =
\frac{F(0)}{F(\infty)} .
\end{equation}
A perturbative calculation based on the results of section~\ref{d>5}
yields
\begin{eqnarray}
\frac{\Phi(z, 0)}{\Phi(z, \infty)} &=& - \frac{g^{2} (8
\pi)^{-(d-1)/2}}{4 (d-3) \rho^{3/2}} \Bigl( \frac{z}{\sqrt{\rho}}
\Bigr)^{\epsilon/2} + O(g^{4}) \nonumber \\
&=& - \frac{u}{6} + O(u^{2}, u \epsilon) .
\end{eqnarray}
Upon application of the renormalization group transformation
with the choice~(\ref{ell}) for the flow parameter this becomes 
\begin{equation}
D 
= - \frac{\epsilon}{9} + O(\epsilon^{2}) .
\end{equation}

In the upper critical dimension $d=5$ the solution of the flow
equation~(\ref{flowu}) reads
\begin{equation}
\bar{u}(l) \simeq \frac{2}{3 \ln(1/l)} \qquad \mbox{for $l \to 0$} .
\end{equation}
This yields for the profile
\begin{equation}
\frac{\Phi(z, 0)}{\Phi(z, \infty)} \simeq - \frac{2}{9 \ln(z/z_{0})}
\qquad \mbox{for $z \to \infty$,}
\end{equation}
where $z_{0}$ is non-universal.

\subsection{Distant wall corrections}

Until now the profile near a particle source has been investigated
assuming that the particles are extracted by a distant sink located
at $z = L$, $L \to \infty$. In computer simulations only comparatively small
systems can by studied and corrections to the profile~(\ref{prof}) due
to the distant sink become important. At mean field level the profile
which satisfies the boundary conditions $\Phi(0) = \infty$ and $\Phi(L) =
-\infty$ is given by
\begin{equation}
\Phi_{\text{mf}}(z) = \frac{2 \pi \rho}{g L} \cot\left(\frac{\pi z}{L}
\right) = \frac{2 \rho}{g z} \Bigl[ 1 - \frac{1}{3} \Bigl( \frac{\pi z}{L}
\Bigr)^{2} + \ldots\Bigr] . \label{SDEmf}
\end{equation}
The powers of $(z/L)$ occurring in this expansion below the upper critical 
dimension can be obtained from a short distance expansion (SDE)
of the order parameter field $s(z)$ for $z \rightarrow 0$~\cite{EKD,EE,D97}.
The leading term in this SDE (with the lowest momentum dimension) is the
unit operator $1$. Since $s_{s}=0$ due to the Dirichlet boundary condition
the next-to-leading contribution is the normal derivative
$\rho\partial_{n} s$. We therefore obtain
\begin{equation}
s({r}_{\bot}, z, t) = A_{1} z^{-\sigma} \cdot 1 + A_{2}
z^{\frac{2-\eta}{2+\eta}} \cdot \rho \partial_{n} s({r}_{\bot}, t) +
\ldots\ . \label{SDE}
\end{equation}
The power in front of the normal derivative has been determined by
comparing the anomalous scaling dimensions of the individual terms in
equation~(\ref{SDE}),
\begin{equation}
s \sim l^{(d-1+\eta)/2} , \qquad \rho\partial_{n} s \sim l^{(d+3-\eta)/2}
, \qquad z \sim l^{-(2+\eta)} .
\end{equation}
The SDE~(\ref{SDE}) implies that the distant particle sink gives rise to
a correction to the profile proportional to $z^{\frac{2-\eta}{2+\eta}} =
z^{\sigma}$ for $z \to 0$, i.e.
\begin{equation}
\Phi(z) = A_{1} z^{-\sigma} \Bigl[ 1 + B \Bigl(\frac{z}{L}\Bigr)^{2\sigma}
+ \ldots \Bigr] . \label{DWC}
\end{equation}
For $\epsilon = 0$ this form is consistent with the mean field 
result~(\ref{SDEmf}).

The amplitudes $A_{1}$ and $B$ depend on the fixed point value of the
surface potential, i.e. they take different values for $h_{1}=0$ and
$h_{1}=\infty$. Equation~(\ref{SDEmf}) shows that for $h_{1}=\infty$ the
(universal) amplitude $B$ is given by $B = -\pi^{2}/3 + O(\epsilon)$.
In the case $h_{1}=0$ with the boundary conditions $\Phi_{\text{mf}}(0)
= 0$ and $\Phi(L) = -\infty$ the mean field profile reads
\begin{equation}
\Phi_{\text{mf}}(z) = -\frac{\pi \rho}{g L} \tan\left(\frac{\pi z}{2 L}
\right) = -\frac{\pi^{2} \rho}{2 g z} \Bigl[ \Bigl( \frac{z}{L}
\Bigr)^{2} + \ldots\Bigr] . \label{SDEmf2}
\end{equation}
To determine the amplitude $B$ at leading order in $\epsilon$ we divide
$\Phi_{\text{mf}}(z)$ by the semi-infinite profile~(\ref{Phi1}) and obtain
\begin{equation}
\frac{\Phi_{\text{mf}}(z)}{\Phi^{[1]}(z)} = \frac{3 \pi^{2}}{2 u} \bigl(
1 + O(u, \epsilon) \bigr) \Bigr(\frac{\mu^{2} z}{\sqrt{\rho}}\Bigl)^{
-\epsilon/2} \Bigl( \frac{z}{L} \Bigr)^{2} + \ldots \ .
\end{equation}
At the fixed point $u_{\star} = (2/3)\epsilon + O(\epsilon^{2})$ the
amplitude $B$ is thus given by $B = 9 \pi^{2}/(4\epsilon) +
O(\epsilon^{0})$. Note that $B$ is of the order $1/\epsilon$ because
the semi-infinite profile vanishes at zero loop order.

\section{Simulation results}

In order to check some of the results presented in the previous section
by computer simulations we use the standard Monte Carlo technique with
Metropolis spin-exchange jump rates on the two-dimensional,
driven Ising lattice gas with attractive interactions~\cite{KLS,K-tL}.
The driving force is effectively infinity, i.e. every attempt of a
particle to jump in the direction of the driving force is successful
unless the jump would violate the excluded volume constraint. Jumps in
the direction antiparallel to the driving force have zero probability.
We use the critical value of the temperature parameter $T_{c}(\infty) =
1.41 T_{c}(0)$ obtained by Leung~\cite{K-tL}. The particle reservoirs at
the boundaries are incorporated into the model by a simple change of the
updating algorithm:
Whenever a boundary site is involved in an updating step the
occupation number of this site set equal to a random number $X \in \{0,
1\}$ which is one with propability $c_{A}$ (at the left boundary) or
$c_{B}$ (at the right boundary). To avoid unwanted correlations each
realization of $X$ has to be used for only one update.
In the transverse directions periodic boundary conditions are imposed.

\begin{figure}
\epsfxsize=330pt
\vspace*{-48mm}\hspace*{-19mm}
\epsfbox{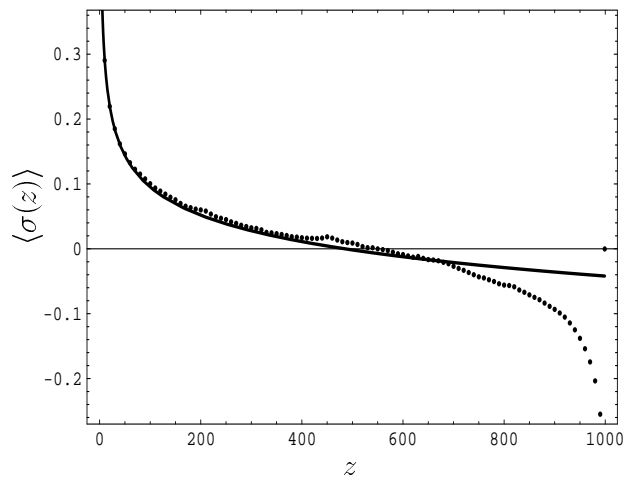}
\vspace*{-47mm}
\caption{Density profile for $c_{A}=1.0$, $c_{B}=0.5$, where the occupation
numbers are represented by the spin variable $\sigma=2n-1$. The statistical
error is everywhere smaller than $\pm 0.006$. The solid line is a fit
using Eq.~(\ref{DWC}) with $A_{1}=0.678$ and $B=-1.62$.}
\label{99-50}
\end{figure}

Fig.~\ref{99-50} shows the density profile for $c_{A}=1.0$ and
$c_{B}=0.5$. The system size is $L_{\|}=1000$ in the direction parallel
to the driving force and $L_{\bot}=500$. At the beginning of each run,
an uncorrelated initial state is generated where each lattice site is
occupied with probability $0.5$. Then $10^{5}$ Monte Carlo steps (per
site) are performed to reach the stationary state. The profile shown in
Fig.~\ref{99-50} has been obtained by averaging over $2\cdot 10^{5}$
configurations. The amplitudes $A_{1}$ and $B$ in Eq.~(\ref{DWC}) have
been determined by a least square fit with the result $A_{1}=0.678 \pm
0.004$, $B = -1.6 \pm 0.2$. For this fit we have used various
subintervals of $3 \leq z \leq 50$. (The statistical error in this
range is smaller than $0.002$.) For larger values of $z$ higher powers
in $z/L$ become increasingly important. We have checked that the above
values for $A_{1}$ and $B$ also provide acceptable fits for smaller
systems [$(L_{\|}, L_{\bot})=(500, 397)$ and $(125, 250)$].

\begin{figure}
\epsfxsize=330pt
\vspace*{-48mm}\hspace*{-19mm}
\epsfbox{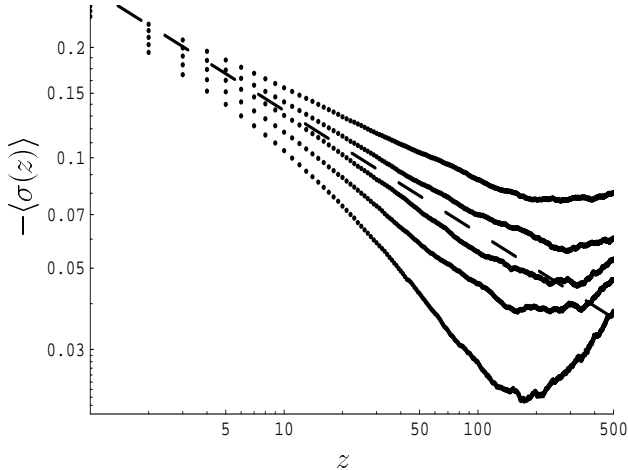}
\vspace*{-47mm}
\caption{Double logarithmic plot of the density profile for
$c_{A}=0.278$, $0.280$, $0.282$, $0.284$, $0.286$ (from top to bottom)
and $c_{B} = 0.5$. The spin variable $\sigma = 2n-1$ has been used.
The broken line corresponds to the power $0.29 z^{-1/3}$.}  
\label{loglog}
\end{figure}
\begin{figure}
\epsfxsize=330pt
\vspace*{-48mm}\hspace*{-19mm}
\epsfbox{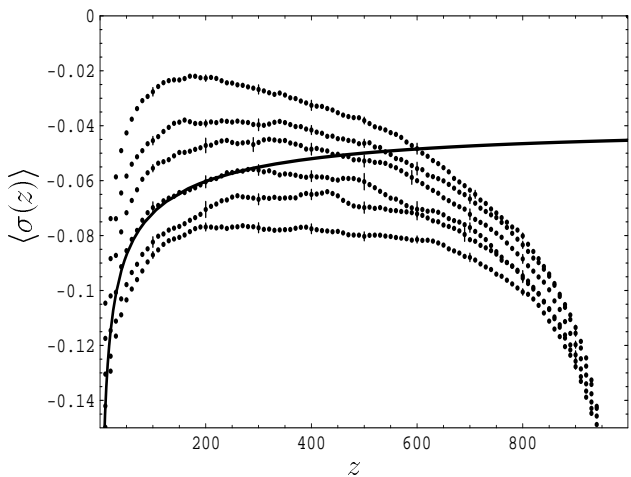}
\vspace*{-47mm}
\caption{The density profile for $c_{A}=0.278$, $0.279$, $0.280$, $0.282$,
$0.284$, $0.286$ (from bottom to top) and $c_{B} = 0.5$. The occupation
numbers are represented by the spin variable $\sigma = 2n-1$. The solid
curve is a fit using Eq.~(\ref{DWC}) with $A_{1}= -0.30$ and $B= 0.51$.}
\label{h1crit}
\end{figure}

To determine the amplitude ratio $D$ one first has to find the
critical value of $c_{A}$ that corresponds to a vanishing
surface field, $h_{1}=0$. Fig.~\ref{loglog} suggests that this value
is close to $c_{A} \approx 0.28$. For $0.278 \leq c_{A} \leq 0.282$
we obtained fits consistent with $A_{1} = -0.29 \pm 0.02$. One of
these fits is depicted in Fig.~\ref{h1crit} together with density
profiles for various values of $c_{A}$. Each profile is an average over
$10^{6}$ configurations.  To determine the amplitude $B$ it would be
necessary to obtain a more accurate estimate for the critical surface
density.
The simulation result for the amplitude ratio reads $D = -0.43
\pm 0.03$ which can be compared with our one-loop calculation, $D
\approx -\epsilon/9 = -0.33$ for $\epsilon=3$.

\section{Summary and outlook}
\label{Disc}

A particle reservoir coupled to the boundary of a driven diffusive
system maintains the critical density in the bulk if the chemical
potential of the reservoir is not below a critical value. Above this
critical value the density profile (as a function of the distance from
the boundary) asymptotically approaches the bulk density from above, where
the decay of the profile follows a power law with an exponent $\sigma$
which can be expressed in terms of the bulk exponent $\eta$.
At the critical value of the boundary chemical potential the density
tends to its bulk value from below. If the chemical potential is close
to (but above) its critical value the density profile crosses the
critical density at a macroscopic distance $\zeta$ from the boundary.
The singular power law dependence of the length scale $\zeta$ on the
boundary chemical potential is characterized by a new exponent
$\nu_{1}$ which has been calculated at first order in $\epsilon = 5-d$.

While in exclusion models without particle-particle attraction the
density profile is always a monotonic function of the distance from the
boundary we have shown that in critical DDS stationary profiles can have
local maximum points. This is due to the density correlations in
the bulk generated (for $d > 1$) by the attractive interaction. If the
`temperature' is raised above $T_{c}(E)$ these correlations survive as
long as $T$ is finite. Therefore the {\em qualitative} form of the
density profile will not change for $T_{c}(E) < T < \infty$.

In this paper one out of a multitude of universality classes
describing various types of DDS has been considered. These universality
classes differ in the nature of the noise (particle conserving or
non-conserving), the presence or absence of quenched disorder and the
values of temperature-like critical parameters~\cite{Volker}. We plan to
extend the analysis presented here to other universality classes.
It is straightforward to derive relations similar to~(\ref{sig})
between $\sigma$ and the anisotropy exponent $\eta$ for DDS with
quenched disorder. This makes it possible to check the field
theoretic predictions of Refs.~\cite{BJ,SB,Volker} for disordered DDS by
Monte Carlo simulations of the density profile in systems with open
boundaries. Note that in the presence of quenched disorder periodic
boundary conditions (in the direction parallel to the driving force) may
lead to unwanted correlations since the particles are subjected to the
same randomness after every passage through the system.

In order to obtain a numerical estimate for the surface exponent
$\nu_{1}$ or a more accurate value for the amplitude ratio $D$ it is
necessary to determine the critical surface density more accurately.
This is an open problem for future simulations.


\acknowledgments
This work has been supported in part by the Sonderforschungsbereich 237
[Unordnung und Gro{\ss}e Fluktuationen (Disorder and Large
Fluctuations)] of the Deutsche Forschungsgemeinschaft.

\appendix

\section{Surface divergences at one-loop order}

\begin{figure}
\begin{center}
\hspace*{3mm}
\epsfxsize=200pt\epsfbox{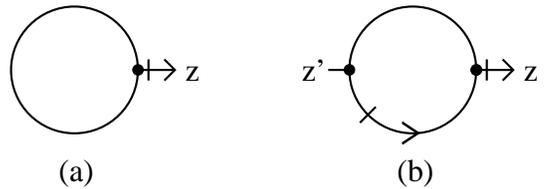}
\vspace*{5mm}
\caption{Ultraviolet divergent Feynman diagrams at one--loop order.
A line with (without) an arrow represents the Gaussian propagator
(correlator). The short line perpendicular to the propagator line in
the diagram (b) indicates a derivative with respect to $z^{\prime}$.}
\label{feynm}
\end{center}
\end{figure}

In order to determine the renormalization constants at one--loop order
one has to evaluate the ultraviolet divergent Feynman diagrams shown
in Fig.~\ref{feynm}. The results have to be interpreted in the
distribution sense since the calculation of Green functions involves
integrations over the $z$-coordinates of amputated graphs.

The Laplace transform of the first diagram in Fig.~\ref{feynm} reads
\begin{eqnarray}
-\frac{\lambda g}{2}\int_{0}^{\infty} dz e^{-s z}
\int_{{\bf q}_{\bot}, \omega} \hat{C}_{{\bf q}_\bot, \omega}(z, z)
= - \frac{\lambda g}{\epsilon} A_{\epsilon} \tau^{-\epsilon/2}
\nonumber \\
\times
\Bigl( \frac{\tau^{2}}{\sqrt{\rho} s} + \frac{4\tau}{3}
+ \frac{2}{3} \sqrt{\rho} s + O(\epsilon) \Bigr) , \label{a1}
\end{eqnarray}
where $\hat{C}_{{\bf q}_\bot, \omega}(z, z)$ is the Gaussian
correlator~(\ref{corr}) at the Dirichlet fixed point.
The two dimensional Laplace transform of the second diagram
is given by
\begin{eqnarray}
(\lambda g)^{2} \int_{0}^{\infty} dz e^{-s z}
\int_{0}^{\infty} dz^{\prime} e^{-s^{\prime} z^{\prime}}
\nonumber \\
\times \int_{{\bf q}_{\bot}, \omega} \hat{C}_{-{\bf q}_\bot, -\omega}(z,
z^{\prime}) \partial_{z^{\prime}} \hat{G}_{{\bf q}_\bot, \omega}(z,
z^{\prime}) \label{a2} \\
= \frac{\lambda g^{2}}{2 \sqrt{\rho} \epsilon} 
A_{\epsilon} \tau^{-\epsilon/2} \Bigl( \frac{2 s^{\prime}}{s+s^{\prime}}
- \frac{2}{3} + O(\epsilon) \Bigr) . \nonumber
\end{eqnarray}  
Applying the inverse Laplace transformation to~(\ref{a1}) and~(\ref{a2})
we obtain
\begin{eqnarray}
\mbox{Graph \ref{feynm}(a)} = - \frac{\lambda g}{\epsilon} A_{\epsilon}
\tau^{-\epsilon/2} \Bigl( \frac{\tau^{2}}{\sqrt{\rho}} + \frac{4\tau}{3}
\delta(z) \nonumber \\
+ \frac{2}{3} \sqrt{\rho} \delta^{\prime}(z) + O(\epsilon) \Bigr)
\label{a3}
\end{eqnarray}
and
\begin{eqnarray}
\mbox{Graph \ref{feynm}(b)} =  \frac{\lambda g^{2}}{2 \sqrt{\rho}
\epsilon} A_{\epsilon} \tau^{-\epsilon/2} \Bigl( 2
\delta^{\prime}(z^{\prime} \mid z) \nonumber \\
- \frac{2}{3} \delta(z^{\prime}) \delta(z) + O(\epsilon) \Bigr) ,
\label{a4}
\end{eqnarray}
where we have introduced the definition
\begin{equation}
\int_{0}^{\infty} dz^{\prime} \delta^{\prime}(z^{\prime}|z) f(z^{\prime})
= -f^{\prime}(z).
\end{equation}

The $\epsilon$-poles are canceled by the counter terms~(\ref{bct})
and~(\ref{sct}) given in section~\ref{renorm}.
The values of the coefficients $A$, $B$, $F$, $K$, and $Z$ at
one-loop order follow from Eqs.~(\ref{a3}) and~(\ref{a4}) as
\begin{equation}
A = -\frac{1}{\epsilon} \qquad B = -\frac{u}{3 \epsilon} \qquad
F = -\frac{4}{3 \epsilon} \qquad K = \frac{2}{3 \epsilon}
\end{equation}
and
\begin{equation}
Z = 1 - \frac{u}{\epsilon} .
\end{equation}



\begin{thebibliography}{0}
\bibitem{footnote}The integral over the gradient $\partial_{\|} \tilde{s}$
(which cancels in a system with periodic boundary conditions) can be written as
$\int_{V} d^{d}r \partial_{\|} \tilde{s} = \int_{\partial V} d^{d-1}r
[\tilde{s}(z \to \infty) - \tilde{s}_{s}]$. Due to particle conservation the
contribution $\tilde{s}(z \to \infty)$ may not be omitted since response
functions with $q_{\bot}=\omega=0$ are non-zero for $z\to\infty$.
\end{thebibliography}
\end{document}